\documentclass[11pt]{article}
\newcommand{\be}{\begin{equation}}
\newcommand{\ee}{\end{equation}}
\newcommand{\bea}{\begin{eqnarray}}
\newcommand{\eea}{\end{eqnarray}}
\newcommand{\sn}{{\rm sn}}

\newcommand{\dn}{{\rm dn}}
\newcommand{\cn}{{\rm cn}}
\newcommand{\sech}{{\rm sech}}

\begin{document}
\vspace{0.5in}

\begin{center}
{\LARGE{\bf New Solutions of Nonlocal NLS, mKdV and Hirota Equations}}
\end{center}

\begin{center}
{\LARGE{\bf Avinash Khare}} \\
{Physics Department, Savitribai Phule Pune University \\
 Pune 411007, India}
\end{center}

\begin{center}
{\LARGE{\bf Avadh Saxena}} \\ 
{Theoretical Division and Center for Nonlinear Studies, 
Los Alamos National Laboratory, Los Alamos, New Mexico 87545, USA}
\end{center}

\begin{abstract}
In this paper, we provide several novel solutions of the 
Ablowitz-Musslimani as well Yang's versions of the nonlocal nonlinear 
Schr\"odinger (NLS) equation, nonlocal modified Korteweg-de Vries 
(mKdV) as well as nonlocal Hirota equations. In each case we compare
and contrast the corresponding solutions of the relevant local equation. 
Further, we provide new solutions of the local NLS, local mKdV and local 
Hirota equations which are not the solutions of the corresponding nonlocal 
equations.
\end{abstract}


\section{Introduction}
 
After the seminal papers of Ablowitz and Musslimani (AM) \cite{am1, am2} 
about the nonlocal NLS equation and its integrability, in recent years 
nonlocal mKdV \cite{mkdv,gur} as well as well as nonlocal Hirota equations 
\cite{cen,hirota} have also been introduced. Moreover, Yang \cite{yang}
has also introduced another variant of the nonlocal NLS equation and shown
its integrability. During the last few years it has been realized that optics and 
photonics can provide an ideal ground for testing some of the consequences 
of the AM variant of the nonlocal NLS \cite{optics}. In general, beyond nonlinear 
optics and photonic waveguides, nonlocal, nonlinear equations arise and find 
applications in a number of other physical contexts \cite{lou} spanning condensed 
matter physics, high energy physics, hydrodynamics, electromagnetics, elasticity, etc. 

Several years ago, we 
obtained \cite{ks1, ks3} a large number of solutions of the AM version of the 
nonlocal NLS.  We might add here that some aspects of the nonlocal mKdV 
equation \cite{am3,am4,li,zh} and nonlocal Hirota equation \cite{li1} 
including few exact solutions have also been reported in the literature. 
Further, recently we have also 
obtained several solutions of the nonlocal mKdV, nonlocal Hirota and Yang 
variant of the nonlocal NLS \cite{ks2}. 
Additionally, in the same paper \cite{ks2} we showed that all these nonlocal 
equations admit superposed solutions which can be re-expressed as a sum of 
two kinks, or kink-anti-kink or two pulse solutions. 

The feeling at that time was that perhaps one has exhausted the various 
possible solutions of the nonlocal 
equations. Of course, being nonlinear equations, one can never be sure about it. 
Recently, we discovered new solutions of the symmetric $\phi^4$ equation 
\cite{kbs}. Subsequently, we realized that even the above mentioned nonlocal 
equations will admit similar novel solutions. The purpose of this paper is to present 
new solutions of the above nonlocal equations. In each case we also enquire if the 
corresponding local equation admits such a solution as well, and if yes, under 
what conditions. Finally, for completeness, in three Appendices we present those 
solutions of the local NLS, mKdV and Hirota equations which are not the solutions 
of the corresponding nonlocal equations.  

The plan of the paper is the following. In Sec. II we present new solutions of the 
nonlocal AM variant of the NLS equation. In each case we
compare and contrast with the solutions of the (local) NLS equation (in 
case they are admitted). In Sec. III we present novel solutions of the
Yang variant of the nonlocal NLS. In Appendix A we present those solutions 
of the local NLS equation which are not the solutions of either the AM or the 
Yang nonlocal variant of the NLS equation. In Sec. IV we present new 
solutions of nonlocal mKdV equation and compare and contrast them with the 
corresponding solutions of the local mKdV. In Appendix B we present 
solutions of the local mKdV which are not the solutions of the nonlocal
mKdV. In Sec. V we present new solutions of nonlocal Hirota equation 
and compare and contrast them with the corresponding solutions of the local 
Hirota equation. In Appendix C we present solutions of the local Hirota equation 
which are not the solutions of the nonlocal Hirota equation.

\section{New Solutions of Ablowitz-Musslimani Variant of Nonlocal NLS 
Model}

The Ablowitz-Musslimani (AM) variant of the nonlocal NLS is given by
\be\label{1}
i\psi_t(x,t) + \psi_{xx}(x,t) + g \psi^2(x,t) \psi^{*}(-x,t) = 0\,,
\ee
where for comparison with local NLS, we have set the coupling constant
to be $g$ instead of $2g$. 
We now show that apart from the several solutions 
already obtained, this equation admits several new solutions, both 
real and complex which we list one by one. We also compare and contrast
the solutions admitted by the corresponding local NLS equation
\be\label{1a}
i\psi_t(x,t) + \psi_{xx}(x,t) + g |\psi|^2 \psi(x,t) = 0\,.
\ee

It is worth reminding that while the local NLS Eq. (\ref{1a}) admits the 
plane wave solution
\be\label{1d}
\psi(x,t) = A e^{i(kx-\omega t)}\,,
\ee
where
\be\label{1e}
\omega = k^2 - gA^2\,,
\ee
the AM variant nonlocal NLS Eq. (\ref{1}) admits a rather unusual plane
wave solution
\be\label{1f}
\psi(x,t) = A e^{(kx-i\omega t)}\,,
\ee
provided
\be\label{1g}
\omega = -k^2 - gA^2\,.
\ee
As a result, while corresponding to any stationary soliton solution of local 
NLS Eq. (\ref{1a}) there is always a moving soliton solution with velocity
$2k$ and $\omega$ being replaced by $\omega - k^2$. However, the 
nonlocal NLS Eq. (\ref{1}) only admits stationary soliton solutions. As an
illustration, local NLS Eq. (\ref{1}) admits the solution
\be\label{1m}
\psi(x,t) = A \sech(\beta(x-vt)) e^{i(kx-\omega t)}\,,
\ee
provided
\be\label{1h}
v = 2k\,,~~\omega = k^2 -\beta^2\,,~~A^2 = 2\beta^2\,.
\ee
On the other hand, nonlocal NLS Eq. (\ref{1}) only admits the solution
\be\label{1j}
\psi(x,t) = A \sech(\beta x) e^{-i\omega t}\,,
\ee
provided
\be\label{1k}
\omega = -\beta^2\,,~~A^2 = 2\beta^2\,.
\ee

We now present 11 new (3 real and 8 complex but PT-invariant) 
solutions of Eq. (\ref{1}) and compare these with those admitted by the 
corresponding local NLS Eq. (\ref{1a}). 
 As we show in the next section, the corresponding Yang's
 nonlocal NLS variant in contrast admits only 3 new solutions. 
 It turns out that there are solutions admitted by the local
 NLS Eq. (\ref{1a}), but not by the nonlocal Eq. (\ref{1}) (or 
 Yang's nonlocal variant), which we present in Appendix A.

{\bf Solution I}

It is easy to check that 
\be\label{2}
\psi(x,t) = e^{i\omega t} \frac{A\sqrt{m}\sn(\beta x,m)}{D+\dn(\beta x,m)}\,,
~~A, D > 0\,,
\ee
is an exact solution of the nonlocal Eq. (\ref{1}) provided $g > 0$. 
\be\label{3}
D = 1\,,~~2 g A^2 = \beta^2\,,~~\omega = -(2-m)\frac{\beta^2}{2}\,.
\ee
Note, here $m$ is the modulus of Jacobi elliptic functions \cite{as}.

In contrast the local NLS Eq. (\ref{1a}) admits the solution (\ref{2})
provided  
\be\label{4}
D = 1\,,~~2 g A^2 = -\beta^2\,,~~\omega = -(2-m)\frac{\beta^2}{2}\,.
\ee
Thus in contrast to the nonlocal case, in the local case the solution is
only valid for $g < 0$.  

{\bf Solution II}

It is easy to check that 
\be\label{5}
\psi(x,t) = e^{i\omega t} \frac{A\sqrt{m}\cn(\beta x,m)}{D+\dn(\beta x,m)}\,,
~~A, D > 0\,,
\ee
is an exact periodic pulse solution of nonlocal Eq. (\ref{1}) provided
\be\label{8}
D^2 = 1-m > 0\,,~~2 g A^2 = -m \beta^2\,,~~\omega = -(2-m)\frac{\beta^2}{2}\,.
\ee
Remarkably the local NLS Eq. (\ref{1a}) also admits the solution (\ref{5})
provided relations (\ref{8}) are satisfied.

{\bf Solution III}

It is easy to check that 
\be\label{9}
\psi(x,t) = e^{i\omega t} \left[F-\frac{A\dn(\beta x,m)}{D+\dn(\beta x,m)}\right]\,,
~~D,F,A > 0\,,
\ee
is an exact periodic pulse solution of the nonlocal Eq. (\ref{1}) provided
\bea\label{10}
&&D^2 = \sqrt{1-m} > 0\,,~~ g A^2 = -2(1-\sqrt{1-m})^2 \beta^2\,,
\nonumber \\
&&F = A/2\,,~~\omega = -[(2-m)+6\sqrt{1-m}]\frac{\beta^2}{2}\,.
\eea
Remarkably the local NLS Eq. (\ref{1a}) also admits the solution (\ref{9})
provided the relations (\ref{10}) are satisfied.

{\bf Complex PT-invariant Periodic and Hyperbolic Pulse 
and Kink Solutions}

We now show that one has several periodic and hyperbolic complex
PT-invariant pulse as well as kink solutions which are distinct from 
the well known  periodic and  hyperbolic complex PT-invariant pulse
and kink solutions. We discuss these one by one. It is worth noting that,
in contrast, the local NLS Eq. (\ref{1a}) does not admit any of the complex
PT-invariant periodic solutions as given below (see Solutions IV to XI).

{\bf Solution IV}

It is easy to check that 
\be\label{11}
\psi(x,t) = e^{i\omega t} \frac{\sqrt{m}[A\cn(\beta x,m)+iB\sn(\beta x,m)]}
{D+\dn(\beta x,m)}\,,
~~A, D > 0\,,
\ee
is an exact complex PT-invariant periodic solution with PT-eigenvalue 
$1$ of the nonlocal Eq. (\ref{1}) provided
\be\label{12}
2g A^2 = (D^2-1)\beta^2\,,~~2g B^2 = (D^2-1+m)\beta^2\,,~~
\omega = -(2-m)\frac{\beta^2}{2}\,.
\ee
Clearly such a solution exists if either
\be\label{13}
D^2 > 1\,,~~g > 0\,,
\ee
or
\be\label{14}
0 < D^2 < 1-m\,,~~g < 0\,.
\ee
Note that for both these cases $\omega < 0$. 

{\bf Solution V}

In the limit $m = 1$, the solution IV goes over to the complex 
PT-invariant hyperbolic solution with PT-eigenvalue $1$, i.e.
\be\label{15}
\psi(x,t) = e^{i\omega t} \frac{[A\sech(\beta x)+iB\tanh(\beta x)]}
{D+\sech(\beta x)}\,,~~A, D > 0\,,
\ee
provided
\be\label{16}
2g A^2 = (D^2-1)\beta^2\,,~~2 g B^2 = D^2\beta^2\,,~~
\omega = -\frac{\beta^2}{2}\,.
\ee
Thus this solution only exists if $g > 0, D^2 > 1$. 

{\bf Solution VI}

It is straightforward to check that 
\be\label{17}
\psi(x,t) = e^{i\omega t} \frac{\sqrt{m}[A\sn(\beta x,m)+iB\cn(\beta x,m)]}
{D+\dn(\beta x,m)}\,,~~A, D > 0\,,
\ee
is an exact complex PT-invariant periodic solution with PT-eigenvalue 
$-1$ of nonlocal Eq. (\ref{1}) provided
\be\label{18}
2g  A^2 = (D^2-1+m)\beta^2\,,~~2g B^2 = (D^2-1)\beta^2\,,~~
\omega = -(2-m)\frac{\beta^2}{2}\,.
\ee
Thus such a solution exists if either
\be\label{19}
D^2 > 1\,,~~g > 0\,,
\ee
or
\be\label{20}
0 < D^2 < 1-m\,,~~g < 0\,.
\ee
Note that for both these cases $\omega < 0$. 

{\bf Solution VII}

In the limit $m = 1$, the solution VI goes over to the complex 
PT-invariant hyperbolic solution with PT-eigenvalue $-1$, i.e.
\be\label{21}
\psi(x,t) = e^{i\omega t} \frac{[A\tanh(\beta x)+iB\sech(\beta x)]}
{D+\sech(\beta x)}\,,~~A, D > 0\,,
\ee
provided
\be\label{22}
2 g A^2 = D^2 \beta^2\,,~~2g B^2 = (D^2-1) \beta^2\,,~~
\omega = -\frac{\beta^2}{2}\,.
\ee
Thus this solution only exists if $g > 0$ and $D^2 > 1$. 

{\bf Solution VIII}

It is easy to check that 
\be\label{23}
\psi(x,t) = e^{i\omega t} \frac{[A\dn(\beta x,m)+iB\sqrt{m}\sn(\beta x,m)]}
{D+\cn(\beta x,m)}\,,~~A > 0, D > 1\,,
\ee
is an exact complex PT-invariant periodic solution with PT-eigenvalue 
$1$ of the nonlocal Eq. (\ref{1}) provided
\be\label{24}
2g A^2 = (D^2-1)\beta^2\,,~~2 m g B^2 = (m D^2-m+1)\beta^2\,,~~
\omega = -(2m-1)\frac{\beta^2}{2}\,.
\ee
Thus this solution exists only if $g > 0$. 

{\bf Solution IX}

It is easy to check that 
\be\label{25}
\psi(x,t) = e^{i\omega t}\frac{[A\sqrt{m}\sn(\beta x,m)+iB\dn(\beta x,m)]}
{D+\cn(\beta x,m)}\,,~~A,B > 0, D > 1\,,
\ee
is an exact complex PT-invariant periodic solution with PT-eigenvalue 
$-1$ of the nonlocal Eq. (\ref{1}) provided
\be\label{26}
2 m g A^2 = (1-m+m D^2)\beta^2\,,~~2 g B^2 = (D^2-1)\beta^2\,,~~
\omega = -(2m-1)\frac{\beta^2}{2}\,.
\ee
Thus this solution exists only if $g > 0$. 

{\bf Solution X}

It is easy to check that 
\be\label{27}
\psi(x,t) = e^{i\omega t}\frac{[A+iB\sin(\beta x)]}
{D+\cos(\beta x)}\,,~~A,B > 0, D > 1\,,
\ee
is an exact complex PT-invariant periodic solution with PT-eigenvalue 
$1$ of the nonlocal Eq. (\ref{1}) provided
\be\label{28}
2 g A^2 = (D^2-1)\beta^2\,,~~2 g B^2 = \beta^2\,,~~
\omega = \frac{\beta^2}{2}\,.
\ee
Thus this solution exists only if $g,\omega > 0$. 

{\bf Solution XI}

It is easy to check that 
\be\label{29}
\psi(x,t) = e^{i\omega t}\frac{[A\sin(\beta x)+iB]}
{D+\cos(\beta x)}\,,~~A,B > 0, D > 1\,,
\ee
is an exact complex PT-invariant periodic solution with PT-eigenvalue 
$-1$ of the nonlocal Eq. (\ref{1}) provided
\be\label{30}
2  g A^2 = \beta^2\,,~~2 g B^2 = (D^2-1)\beta^2\,,~~
\omega = \frac{\beta^2}{2}\,.
\ee
Thus this solution exists only if $g,\omega > 0$. 
As mentioned above, 6 solutions of (local) NLS Eq. (\ref{1a}) which 
are not the solutions of either the AM or Yang variant of nonlocal NLS 
Eq. (\ref{1}), are presented in Appendix A.

\section{New Solutions of Yang's Nonlocal NLS Equation}

The Yang variant of the nonlocal NLS is given by
\be\label{4.1}
i\psi_t(x,t) + \psi_{xx}(x,t) + \frac{g}{2}[|\psi(x,t)|^2 +|\psi(-x,t)|^2] 
\psi(x,t) = 0\,,
\ee
where for comparison with local NLS, we have set the coupling constant
to be $g/2$ instead of $g$. 

It is worth reminding that similar to the local NLS, the Yang variant of the
nonlocal NLS equation also admits the plane wave solution (\ref{1d}) where the
dispersion relation is again given by Eq. (\ref{1e}). 
However, while corresponding to any stationary soliton solution of local 
NLS Eq. (\ref{1a}) there is always a moving soliton solution with velocity
$2k$ and $\omega$ being replaced by $\omega -k^2$ (e.g. see Eq. (\ref{1m})), 
the corresponding Yang (as well as AM nonlocal variant) 
equation only admit the stationary soliton solution. 

We now show that apart from the several solutions 
already obtained \cite{ks2}, this equation admits three new solutions.

{\bf Solution I}

It is easy to check that 
\be\label{4.2}
\psi(x,t) = e^{i\omega t} \frac{A\sqrt{m}\sn(\beta x,m)}{D+\dn(\beta x,m)}\,,
~~A, D > 0\,,
\ee
is an exact solution of the nonlocal Eq. (\ref{4.1}) provided $g > 0$
\be\label{4.3}
D = 1\,,~~2 g A^2 = -\beta^2\,,~~\omega = -(2-m)\frac{\beta^2}{2}\,.
\ee

Thus while local NLS and Yang's nonlocal variant admit solution (\ref{4.2})
for $g < 0$, the AM nonlocal variant admits this solution in case $g > 0$.

{\bf Solution II}

It is easy to check that 
\be\label{4.4}
\psi(x,t) = e^{i\omega t} \frac{A\sqrt{m}\cn(\beta x,m)}{D+\dn(\beta x,m)}\,,
~~A, D > 0\,,
\ee
is an exact periodic pulse solution of the nonlocal Eq. (\ref{4.1}) provided
\be\label{4.5}
D^2 = 1-m > 0\,,~~2 g A^2 = -m \beta^2\,,~~\omega = -(2-m)\frac{\beta^2}{2}\,.
\ee
Remarkably the local NLS Eq. (\ref{1a}) as well as AM nonlocal NLS 
variants also admit the solution (\ref{4.4}) provided the relations 
(\ref{4.5}) are satisfied.

{\bf Solution III}

It is easy to check that 
\be\label{4.6}
\psi(x,t) = e^{i\omega t} [F-\frac{A\dn(\beta x,m)}{D+\dn(\beta x,m)}]\,,
~~D,F,A > 0\,,
\ee
is an exact periodic pulse solution of the nonlocal Eq. (\ref{4.1}) provided
\bea\label{4.7}
&&D^2 = \sqrt{1-m} > 0\,,~~ g A^2 = -2(1-\sqrt{1-m})^2 \beta^2\,,
\nonumber \\
&&F = A/2\,,~~\omega = -[(2-m)+6\sqrt{1-m}]\frac{\beta^2}{2}\,.
\eea
The local NLS Eq. (\ref{1a}) as well as the AM nonlocal NLS 
variant also admit the solution (\ref{4.6})
provided the relations (\ref{4.7}) are satisfied.

\section{New Solutions of a nonlocal mKdV Equation}

Recently a nonlocal mKdV equation has been introduced \cite{mkdv}
\be\label{1.1}
u_t(x,t) +u_{xxx}(x,t)+6g u(x,t) u(-x,-t) u_x(x,t) = 0\,.
\ee
Here $g = +1(-1)$ corresponds attractive (repulsive) mKdV.
We now show that apart from the several solutions obtained recently 
\cite{ks2}, there are 26 new solutions of the nonlocal mKdV
Eq. (\ref{1.1}). We also enquire if the corresponding local mKdV 
equation
\be\label{1.1a}
u_t(x,t) +u_{xxx}(x,t)+6g u^2(x,t) u_x(x,t) = 0\,,
\ee
also admits such solutions and if yes under what conditions. For
completeness in Appendix B we mention those solutions which are 
solutions of the local mKdV Eq. (\ref{1.1a}) but not of the nonlocal
Eq. (\ref{1.1}).

{\bf Solution I}

Inspired by the solution of NLS as obtained long time ago by Zakharav
and Shabat \cite{zs}, we now show that the nonlocal mKdV Eq. (\ref{1.1})
admits a similar solution. In particular, it is straightforward to
check that
\be\label{1.1b}
u(x,t) = \sqrt{n}[B\tanh(\xi)+iA]e^{i(kx-\omega t)}\,,~~\xi = 
\beta(x-vt)\,,
\ee
is an exact PT-invariant solution of the nonlocal mKdV Eq. (\ref{1.1}) 
with PT-eigenvalue $-1$ provided
\be\label{1.1c}
A^2+B^2 = 1\,,~~g = +1\,,~~\beta = \sqrt{n} B\,,
\ee
\be\label{1.1d}
\omega = -(6n+k^2)k\,,~~v = 4nB^2-3k^2-6\sqrt{n} Ak-6n\,.
\ee
It is worth pointing out that in contrast the (local) mKdV equation
(\ref{1.1a}) does not admit the solution  (\ref{1.1b}).

{\bf Solution II}

Yet another PT-invariant solution with PT-eigenvalue +$1$ of the 
nonlocal mKdV Eq. (\ref{1.1}) is
\be\label{1.1e}
u(x,t) = \sqrt{n}[A+iB\tanh(\xi)]e^{i(kx-\omega t)}\,,~~\xi = 
\beta(x-vt)\,,
\ee
provided
\be\label{1.1f}
A^2+B^2 = 1\,,~~g = -1\,,~~\beta = \sqrt{n} B\,,
\ee
\be\label{1.1g}
\omega = -(6n+k^2)k\,,~~v = 4nB^2-3k^2+6\sqrt{n} Ak-6n\,.
\ee
It is interesting that while solution I with PT-eigenvalue $-1$ is admitted 
in case $g = 1$, the solution II with PT-eigenvalue +$1$ is admitted only
if $g = -1$. Note that in contrast the (local) mKdV equation (\ref{1.1a})
does not admit the solution (\ref{1.1e}).

{\bf Solution III}

Remarkably, unlike the local mKdV Eq. (\ref{1.1a}), the nonlocal mKdV 
Eq. (\ref{1.1}) admits the plane wave solution
\be\label{7.1}
u(x,t) = A e^{i(kx-\omega t)}\,,
\ee
provided
\be\label{7.2}
\omega = k(6g A^2-k^2)\,.
\ee

It turns out that unlike the local mKdV Eq. (\ref{1.1a}), the nonlocal
mKdV Eq. (\ref{1.1}) admits several real solutions multiplied by the 
plane wave as given by Eq. (\ref{7.1}). We now present 12 such new 
solutions.

{\bf Solution IV}

It is easy to check that the nonlocal mKdV Eq. (\ref{1.1}) admits the
complex periodic kink solution
\be\label{7.3}
u(x,t) = A \sqrt{m} \sn(\xi,m) e^{i(kx-\omega t)}\,,~~\xi = \beta(x-vt)\,,
\ee
provided
\be\label{7.4}
g = 1\,,~~A = \beta\,,~~\omega = -k[k^2+3(1+m)A^2]\,,~~v = -[3k^2
+(1+m)A^2]\,.
\ee
In the limit $\omega = k = 0$ we get back the well known real periodic 
kink solution $u(x,t)=A\sqrt{m} \sn(\xi,m)$ \cite{ks1,ks3}. 

{\bf Solution V}

In the limit $m = 1$, the solution IV goes over to the complex hyperbolic 
kink solution
\be\label{7.5}
u(x,t) = A \tanh(\xi) e^{i(kx-\omega t)}\,,~~\xi = \beta(x-vt)\,,
\ee
provided
\be\label{7.6}
g = 1\,,~~A = \beta\,,~~\omega = -k[k^2+6 A^2]\,,~~v = -[3k^2
+2 A^2]\,.
\ee
In the limit $\omega = k = 0$ we get back the well known real hyperbolic 
kink solution $u(x,t)=A\tanh(\xi)$ \cite{ks1,ks3}. 

{\bf Solution VI}

It is easy to check that the nonlocal mKdV Eq. (\ref{1.1}) admits the
complex periodic pulse solution
\be\label{7.7}
u(x,t) = A \dn(\xi,m) e^{i(kx-\omega t)}\,,~~\xi = \beta(x-vt)\,,
\ee
provided
\be\label{7.8}
g = 1\,,~~A = \beta\,,~~\omega = -k[k^2-3(2-m)A^2]\,,~~v = -[3k^2
-(2-m)A^2]\,.
\ee
In the limit $\omega = k = 0$ we get back the well known real periodic 
pulse solution $u(x,t)=A\dn(\xi,m)$ \cite{ks1,ks3}. 

{\bf Solution VII}

It is straightforward to check that the nonlocal mKdV Eq. (\ref{1.1}) admits another
complex periodic pulse solution
\be\label{7.9}
u(x,t) = A \sqrt{m} \cn(\xi,m) e^{i(kx-\omega t)}\,,~~\xi = \beta(x-vt)\,,
\ee
provided
\be\label{7.10}
g = 1\,,~~A = \beta\,,~~\omega = -k[k^2-3(2m-1)A^2]\,,~~v = -[3k^2
-(2m-1)A^2]\,.
\ee
In the limit $\omega = k = 0$ we get back the well known real periodic 
pulse solution $u(x,t)=A\sqrt{m} \cn(\xi,m)$ \cite{ks1,ks3}. 

{\bf Solution VIII}

In the limit $m = 1$, the solutions VI and VII go over to the complex 
hyperbolic pulse solution
\be\label{7.11}
u(x,t) = A \sech(\xi) e^{i(kx-\omega t)}\,,~~\xi = \beta(x-vt)\,,
\ee
provided
\be\label{7.12}
g = 1\,,~~A = \beta\,,~~\omega = -k[k^2-3 A^2]\,,~~v = -[3k^2
-A^2]\,.
\ee
In the limit $\omega = k = 0$ we get back the well known real hyperbolic 
pulse solution $u(x,t)=A\sech(\xi)$ \cite{ks1,ks3}.

{\bf Solution IX}

It is easy to check that the nonlocal mKdV Eq. (\ref{1.1}) admits the
complex periodic solution
\be\label{7.13}
u(x,t) = \frac{A \sqrt{m} \cn(\xi,m)}{\dn(\xi,m)} e^{i(kx-\omega t)}\,,
~~\xi = \beta(x-vt)\,,
\ee
provided
\be\label{7.14}
g = -1\,,~~0 < m < 1\,,~~A = \beta\,,~~\omega = -k[k^2+3(1+m)A^2]\,,
~~v = -[3k^2+(1+m)A^2]\,.
\ee
In the limit $\omega = k = 0$ we get back the well known real periodic 
solution $u(x,t)=A\sqrt{m} {\cn(\xi,m)}/{\dn(\xi,m)}$ \cite{ks1,ks3}. 

{\bf Solution X}

It is easy to check that the nonlocal mKdV Eq. (\ref{1.1}) admits the
complex periodic solution
\be\label{7.15}
u(x,t) = \frac{A \sqrt{m(1-m)} \sn(\xi,m)}{\dn(\xi,m)} e^{i(kx-\omega t)}\,,
~~\xi = \beta(x-vt)\,,
\ee
provided
\be\label{7.16}
g = -1\,,~~0 < m < 1\,,~~A = \beta\,,~~\omega = -k[k^2-3(2m-1)A^2]\,,
~~v = -[3k^2-(2m-1)A^2]\,.
\ee
In the limit $\omega = k = 0$ we get back the well known real periodic 
solution $u(x,t)={A\sqrt{m} \sn(\xi,m)}/{\dn(\xi,m)}$ \cite{ks1,ks3}. 

{\bf Solution XI}

It is easy to check that the nonlocal mKdV Eq. (\ref{1.1}) admits the
complex periodic solution
\be\label{7.17}
u(x,t) = \frac{A \sqrt{1-m}}{\dn(\xi,m)} e^{i(kx-\omega t)}\,,
~~\xi = \beta(x-vt)\,,
\ee
provided
\be\label{7.18}
g = 1\,,~~0 < m < 1\,,~~A = \beta\,,~~\omega = -k[k^2-3(2-m)A^2]\,,
~~v = -[3k^2-(2-m)A^2]\,.
\ee
In the limit $\omega = k = 0$ we get back the well known real periodic 
solution $u(x,t)=A\sqrt{1-m}/\dn(\xi,m)$ \cite{ks1,ks3}. 

{\bf Solution XII}

It is easy to check that the nonlocal mKdV Eq. (\ref{1.1}) admits the
complex periodic superposed solution
\be\label{7.19}
u(x,t) = [A \dn(\xi,m)+\sqrt{m} B \cn(\xi,m)] e^{i(kx-\omega t)}\,,
~~\xi = \beta(x-vt)\,,
\ee
provided
\bea\label{7.20}
&&g = 1\,,~~0 < m < 1\,,~~B = \pm A\,,~~2A = \beta\,, \nonumber \\
&&\omega = -k[k^2-6(1+m)A^2]\,,~~v = -[3k^2-2(1+m)A^2]\,.
\eea
In the limit $\omega = k = 0$ we get back the well known real periodic 
superposed solution $u(x,t)=A\dn(\xi,m)+B\sqrt{m}\cn(\xi,m)$ \cite{ks1,ks3}. 

{\bf Solution XIII}

It is easy to check that the nonlocal mKdV Eq. (\ref{1.1}) admits the
complex periodic superposed solution
\be\label{7.21}
u(x,t) = \left[A \dn(\xi,m)+\frac{B \sqrt{1-m}}{\dn(\xi,m)}\right] e^{i(kx-\omega t)}\,,
~~\xi = \beta(x-vt)\,,
\ee
provided
\bea\label{7.22}
&&g = 1\,,~~B = \pm A\,,~~A = \beta\,,
~~v = -3k^2 +[2-m \mp 6\sqrt{1-m}]A^2\,, \nonumber \\
&&0 < m < 1\,,~~\omega = -k^3 +3k [2-m \pm 6\sqrt{1-m} A^2]\,.
\eea
In the limit $\omega = k = 0$ we get back the well known real periodic 
superposed solution $u(x,t)=A\dn(\xi,m)+ {B\sqrt{1-m}}/{\dn(\xi,m)}$ 
\cite{ks1,ks3}. 

{\bf Solution XIV}

It is easy to check that the nonlocal mKdV Eq. (\ref{1.1}) admits the
complex periodic superposed solution
\be\label{7.23}
u(x,t) = \left[\frac{A\sqrt{1-m}}{\dn(\xi,m)}+\frac{iB \sqrt{m}\cn(\xi,m)}
{\dn(\xi,m)}\right] e^{i(kx-\omega t)}\,,~~\xi = \beta(x-vt)\,,
\ee
provided
\bea\label{7.24}
&&g = 1\,,~~B = \pm A\,,~~2A = \beta\,,~~\omega 
= -k[k^2+6 (2m-1)A^2]\,, \nonumber \\
&&0 < m < 1\,,~~v = -[3k^2 +2(2m-1)A^2]\,.
\eea
In the limit $\omega = k = 0$ we obtain the complex periodic 
superposed solution 
\be\label{7.25}
u(x,t)=\frac{A\sqrt{1-m}}{\dn(\xi,m)}+\frac{iB\sqrt{m}\cn(\xi,m)}{\dn(\xi,m)}\,,
\ee
provided
\be\label{7.26}
g = 1\,,~~0 < m < 1\,,~~B = \pm A\,,~~2A = \beta\,,
~~v = -2(2m-1)A^2\,.
\ee

{\bf Solution XV}

It is not difficult to check that the nonlocal mKdV Eq. (\ref{1.1}) admits the
complex periodic superposed solution
\be\label{7.27}
u(x,t) = \left[\frac{A\sqrt{m}\cn(\xi,m)}{\dn(\xi,m)}+\frac{iB \sqrt{1-m}}
{\dn(\xi,m)}\right] e^{i(kx-\omega t)}\,,~~\xi = \beta(x-vt)\,,
\ee
provided
\bea\label{7.28}
&&g = -1\,,~~B = \pm A\,,~~2A = \beta\,,~~\omega 
= -k[k^2+6 (2m-1)A^2]\,, \nonumber \\
&&0 < m < 1\,,~~v = -[3k^2 +2(2m-1)A^2]\,.
\eea
In the limit $\omega = k = 0$ we obtain the complex periodic 
superposed solution 
\be\label{7.29}
u(x,t) = \frac{A\sqrt{m}\cn(\xi,m)}{\dn(\xi,m)}+\frac{iB \sqrt{1-m}}
{\dn(\xi,m)}\,,
\ee
provided
\be\label{7.30}
g = -1\,,~~0 < m < 1\,,~~B = \pm A\,,~~2A = \beta\,,
~~v = -2(2m-1)A^2\,.
\ee

{\bf Solution XVI}

Remarkably, unlike local mKdV Eq. (\ref{1.1a}), the nonlocal mKdV 
Eq. (\ref{1.1}) admits yet another plane wave type solution
\be\label{7.31}
u(x,t) = A e^{kx-\omega t}\,,
\ee
provided 
\be\label{7.32}
\omega = k(k^2 +6g A^2)\,.
\ee

{\bf Solution XVII} 

It is straightforward to check that the nonlocal mKdV Eq. (\ref{1.1})
will admit solutions similar to solutions IV to XV except the factor of 
$e^{i(kx-\omega t)}$ is replaced by the factor of $e^{kx-\omega t}$ provided
one replaces $k^3$ by $-k^3$ in the expression for $\omega$ and replace 
$k^2$ by $-k^2$ in the expression for $v$. As an illustration, it is easy to 
check that the nonlocal mKdV Eq. (\ref{1.1}) admits the solution
\be\label{7.33}
u(x,t) = A \sqrt{m} \sn(\xi,m) e^{(kx-\omega t)}\,,~~\xi = \beta(x-vt)\,,
\ee
provided
\be\label{7.34}
g = 1\,,~~A = \beta\,,~~\omega = k[k^2-3(1+m)A^2]\,,~~v = [3k^2
-(1+m)A^2]\,.
\ee
In the limit $\omega = k = 0$ we get back the well known real periodic 
kink solution $u(x,t)=A\sqrt{m} \sn(\xi,m)$ \cite{ks1,ks3}. 

{\bf Solution XVIII}

In order to obtain the other solutions of the nonlocal mKdV Eq. (\ref{1.1}), we 
define a new variable $\xi = \beta (x-vt)$, in terms of 
which the nonlocal mKdV Eq. (\ref{1.1}) takes the form
\be\label{1.2}
\beta^2 u_{\xi \xi \xi}(\xi) = v u_{\xi}(\xi) - 6g u(\xi)u(-\xi) u_{\xi}(\xi)\,.
\ee
It is then clear that those solutions of the (local) mKdV equation
(\ref{1.1a}) for which $u(-\xi) = \pm u(\xi)$ are clearly
also the solution of the nonlocal mKdV Eq. (\ref{1.1}) with same or
opposite sign of $g$. We now present several new solutions of
nonlocal mKdV.

It is not difficult to check that 
\be\label{1.3}
u(\xi) = \frac{A\sqrt{m}\sn(\xi,m)}{D+\dn(\xi,m)}\,,~~D > 0\,,
\ee
is an exact periodic kink solution of the nonlocal mKdV Eq. (\ref{1.1}) 
provided
\be\label{1.4}
g = 1\,,~~D = 1\,,~~4 A^2 = \beta^2\,,~~v = -(2-m)\frac{\beta^2}{2}\,.
\ee
Note that Eq. (\ref{1.3}) is also a solution of the local mKdV 
Eq. (\ref{1.1a}) except $g = -1$ in that case (but otherwise same relations
as given by Eq. (\ref{1.4}).

{\bf Solution XIX}

It is straightforward to check that 
\be\label{1.5}
u(\xi) = \frac{A\sqrt{m}\cn(\xi,m)}{D+\dn(\xi,m)}\,,~~D > 0\,,
\ee
is an exact periodic pulse solution of the nonlocal mKdV Eq. (\ref{1.1}) 
provided
\be\label{1.6}
g = -1\,,~~D^2 = 1-m > 0\,,~~4 A^2 = m \beta^2\,,~~v 
= -(2-m)\frac{\beta^2}{2}\,.
\ee
Note that Eq. (\ref{1.5}) is also a solution of the local mKdV 
Eq. (\ref{1.1a}) provided the relations (\ref{1.6}) are satisfied.

{\bf Solution XX}

It is easy to check that
\be\label{1.7}
u(\xi) = F + \frac{A}{D+\dn(\xi,m)}\,,~~A, D > 0\,,
\ee
is an exact solution of the nonlocal mKdV Eq. (\ref{1.1}) provided
\bea\label{1.8}
&&6g F^2-v = [6D^2 -(2-m)]\beta^2\,,~~2g FA =D[2-m-2D^2]\beta^2\,,
\nonumber \\
&&g A^2 = (D^2-1)(D^2+m-1) \beta^2\,.
\eea
Thus if $g > 0$ then either $D^2 > 1$ or $0 < D^2 < 1-m$. On the other
hand if $g < 0$ then $1-m < D < 1$. 

Note that Eq. (\ref{1.7}) is also a solution of the local mKdV 
Eq. (\ref{1.1a}) provided the relations (\ref{1.8}) are satisfied.

We thus have 4 possibilities:

1. $g = 1\,,~~F> 0\,,~~D^2 < 1-m\,,~~6 F^2 -v < (4-5m)\beta^2$.

2. $g = 1\,,~~F < 0\,,~~D^2 > 1\,,~~6F^2 -v > (4+m)\beta^2$.

3. $g = -1\,,~~F > 0\,,~~(2-m)/2 < D^2 < 1\,,~~-(4+m)\beta^2 < v+6F^2 
< -(2-m) \beta^2$.

4. $g = -1\,,~~F < 0\,,~~1-m < D^2 < (2-m)/2\,,~~-(2-m)\beta^2 < v+6F^2 
< (5m-4) \beta^2$.

Observe that the solution (\ref{1.7}) can be re-expressed as
\be\label{1.9}
u(\xi) = \frac{FD+A+F\dn(\xi,m)}{D+\dn(\xi,m)}\,,
\ee
which oscillates between $u = \frac{A+FD+F}{D+1}$ and 
$\frac{A+FD+F\sqrt{1-m}}{D+\sqrt{1-m}}$. 

There are two special cases in
which the solution XX takes a somewhat simpler form.

{\bf Case a: $F = 0$}

In case $F =0$, the solution (\ref{1.7}) takes the form
\be\label{1.10}
u(\xi) = \frac{A}{D+\dn(\xi,m)}\,,~~A, D > 0\,,
\ee
which holds good provided
\be\label{1.11}
g = -1\,,~~v = -2(2-m)\beta^2\,,~~D^2 = \frac{2-m}{2}\,,~~
4 A^2 = m^2 \beta^2\,.
\ee

{\bf Case b: $A = -FD$}

In case $A = -FD$, the solution (\ref{1.7}) takes the form
\be\label{1.12}
u(\xi) = \frac{F\dn(\xi,m)}{D+\dn(\xi,m)}\,,~~F, D > 0\,,
\ee
which holds good provided
\be\label{1.13}
g = -1\,,~~v = -2(2-m)\beta^2\,,~~D^2 = \frac{2(1-m)}{2-m}\,,~~
 F^2 = \frac{m^2}{2(2-m)} \beta^2\,.
\ee

{\bf Solution XXI}

In the $m = 1$ limit, the solution (\ref{1.7}) goes over to the 
hyperbolic pulse solution
\be\label{1.14}
u(\xi) = F+\frac{A}{D+\sech(\xi)}\,,~~A, D > 0\,,
\ee
provided
\be\label{1.15}
v = \frac{2D^2+1}{2(D^2-1)}\beta^2\,,~~g A^2 = D^2(D^2-1)\beta^2\,,~~
g F^2 = \frac{(2D^2-1)^2}{4(D^2-1)}\beta^2\,.
\ee
Thus $g > (<)$ 0 depending on if $D > (<)$ 1. 
We thus have three possibilities:

1. $g = 1\,,~~F > 0\,,~~D^2 > 1\,,~~6 F^2 -v > 5\beta^2$.

3. $g = -1\,,~~F > 0\,,~~0 < D^2 < 1/2\,,~~-2\beta^2 < v+6F^2 < \beta^2$.

4. $g = -1\,,~~F < 0\,,~~1/2 < D^2 < 1\,,~~-5\beta^2 < v+6F^2 < -2\beta^2$.

Note that Eq. (\ref{1.14}) is also a solution of the local mKdV 
Eq. (\ref{1.1a}) provided the relations (\ref{1.15}) are satisfied.

It is easy to check that this solution does not exist in case 
$FD = -A$. However, it does exist in the other limit of $F =0$.
In case $F =0$, the solution (\ref{1.14}) takes the form
\be\label{1.16}
u(\xi) = \frac{A}{D+\sech(\xi)}\,,~~A, D > 0\,,
\ee
which holds good provided
\be\label{1.17}
g = -1\,,~~v = -2\beta^2\,,~~D^2 = \frac{1}{2}\,,~~
A^2 = \frac{\beta^2}{4}\,.
\ee

{\bf Solution XXII}

It is easy to check that
\be\label{1.18}
u(\xi) = F - \frac{A}{D+\cn(\xi,m)}\,,~~A > 0, D > 1\,,
\ee
is an exact solution of the nonlocal mKdV Eq. (\ref{1.1}) provided
\bea\label{1.19}
&&g = 1\,,~~6F^2-v = [6m D^2-(2m-1)]\beta^2\,, \nonumber \\
&&2FA =D[2m D^2-(2m-1)]\beta^2\,,~~A^2 = (D^2-1)(m D^2+1-m) \beta^2\,. ~~~~~~~
\eea
Note that Eq. (\ref{1.18}) is also a solution of the local mKdV 
Eq. (\ref{1.1a}) provided the relations (\ref{1.19}) are satisfied.

Thus this solution only exists if $g > 0, D^2 > 1\,,~~6F^2-v > (4m+1) \beta^2$.
Note that the solution (\ref{1.18}) can be re-expressed as
\be\label{1.20}
u(\xi) = \frac{FD-A+F\cn(\xi,m)}{D+\cn(\xi,m)}\,,~~A > 0, D > 1\,.
\ee

Unlike the solution XX, this solution does not exist in case $F = 0$ since
here $D > 1$. However, it does exist in case $FD = A$.
In case $A = FD$, the solution (\ref{1.18}) takes the form
\be\label{1.21}
u(\xi) = \frac{F\cn(\xi,m)}{D+\cn(\xi,m)}\,,~~F, D > 0\,,
\ee
which holds good provided
\be\label{1.22}
g = 1\,,~~v = 2(1-2m)\beta^2 > 0\,,~~D^2 = \frac{2(1-m)}{(1-2m)}\,,~~
2 A^2 = \frac{1}{(1-2m)} \beta^2\,.
\ee
Note that this solution exists only if $m < 1/2$. 

{\bf Solution XXIII}

In the limit $m = 0$, the solution  XXII goes over
to the trigonometric solution
\be\label{1.23}
u(\xi) = F - \frac{A}{D+\cos(\xi)}\,,~~A > 0, D > 1\,,
\ee
provided
\be\label{1.24}
g = 1\,,~~6F^2-v = \beta^2\,,~~2FA =D \beta^2\,,
A^2 = (D^2-1) \beta^2\,.
\ee
Thus for this solution $g > 0\,,~~F > 0$.
Note that Eq. (\ref{1.23}) is also a solution of the local mKdV 
Eq. (\ref{1.1a}) provided the relations (\ref{1.24}) are satisfied.

The solution (\ref{1.23}) takes a simpler form in case $FD = A$.
In case $A = FD$, the solution (\ref{1.23}) takes the form
\be\label{1.25}
u(\xi) = \frac{F\cos(\xi)}{D+\cos(\xi)}\,,~~F, D > 0\,,
\ee
which holds good provided
\be\label{1.26}
g = 1\,,~~v = 2\beta^2 > 0\,,~~D^2 = 2\,,~~
F^2 = \frac{\beta^2}{2}\,.
\ee

{\bf Solution XXIV}

It is easy to check that
\be\label{1.27}
u(\xi) = F -\frac{A}{D+\xi^2}\,,~~D > 0\,,
\ee
is an exact solution of the nonlocal mKdV Eq. (\ref{1}) provided 
\be\label{1.28}
g = 1\,,~~v = 3/2D > 0\,,~~A^2 = 4D\,,~~F = \frac{1}{2\sqrt{D}}\,.
\ee
Thus this solution can also be re-expressed as
\be\label{1.27a}
u(\xi) = \frac{\xi^2-3D}{2\sqrt{D}(D+\xi^2)}\,,~~D > 0\,. 
\ee
Note that Eq. (\ref{1.27}) is also a solution of the local mKdV 
Eq. (\ref{1.1a}) provided the relations (\ref{1.28}) are satisfied.
Note that unlike all other solutions discussed so far, this is a 
solution with a power law tail. 

{\bf Solution XXV}

It is straightforward to check that
\be\label{1.31}
u(\xi) = \frac{[A\sqrt{1-m}+iB\sqrt{m}\cn(\beta \xi,m)]}{\dn(\beta \xi,m)}\,,
\ee
is an exact solution of the nonlocal mKdV Eq. (\ref{1}) provided $0 < m < 1$ 
and 
\be\label{1.32}
g = 1\,,~~v = -(2m-1)\frac{\beta^2}{2}\,,~~B = \pm A\,,~~A^2 = 4\beta^2\,.
\ee
Note that Eq. (\ref{1.31}) is also a solution of the local mKdV 
Eq. (\ref{1.1a}) provided the relations (\ref{1.32}) are satisfied.

{\bf Solution XXVI}

It is easy to check that
\be\label{1.33}
u(\xi) = \frac{[A\sqrt{m}\cn(\beta \xi,m)+iB]}{\dn(\beta \xi,m)}\,,
\ee
is an exact solution of the nonlocal mKdV Eq. (\ref{1}) provided $0 < m < 1$ 
and 
\be\label{1.34}
g = -1\,,~~v = -(2m-1)\frac{\beta^2}{2}\,,~~B = \pm A\,,~~A^2 = 4\beta^2\,.
\ee
Note that Eq. (\ref{1.33}) is also a solution of the local mKdV 
Eq. (\ref{1.1a}) provided the relations (\ref{1.34}) are satisfied.

In Appendix B we present 14 solutions of local mKdV Eq. (\ref{1.1a})
which, however, are not the solutions of the nonlocal mKdV Eq. (\ref{1.1}). 

\section{New Solutions of the Nonlocal Hirota Equation}

The nonlocal Hirota equation is given by \cite{hirota}
\be\label{2.1}
iu_t(x,t)+\alpha[u_{xx}(x,t)+2g u^2(x,t) u(-x,-t)] 
+i\beta[u_{xxx}(x,t)+6g u(x,t) u(-x,-t)u_{x}(x,t)] = 0\,.
\ee

We now obtain 19 new solutions of the nonlocal Hirota Eq. (\ref{2.1})
and compare and contrast them with the corresponding solutions of the
local Hirota equation

\be\label{2.1a}
iu_t(x,t)+\alpha[u_{xx}(x,t)+2g |u|^2(x,t) u(x,t)] 
+i\beta[u_{xxx}(x,t)+6g |u|^2(x,t) u_{x}(x,t)] = 0\,,
\ee

{\bf Solution I}

We now show that like the NLS equation \cite{zs} even nonlocal Hirota Eq. (\ref{2.1}) 
admits the PT-invariant solution with PT-eigenvalue $-1$, i.e. 
\be\label{2.19}
u(x,t) = \sqrt{n}\, [B\tanh(\xi)+iA] e^{i(\omega t-kx)}\,,~~\xi = 
\delta(x-vt)\,,
\ee
provided
\bea\label{2.20}
&&g = 1\,,~~A^2+B^2 = 1\,,~~\delta = \sqrt{n} B\,,~~\omega 
= -\alpha(k^2+2n)+\beta k(k^2+6n)\,, \nonumber \\
&&v = -2\alpha(k-\sqrt{n} A) +\beta(4n B^2-3k^2-6n+6kA\sqrt{n})\,.
\eea
Note that Eq. (\ref{2.19}) is also the solution of the 
corresponding local Hirota Eq. (\ref{2.1a}) 
provided unlike the solution I, $g = -1$ in the local Hirota case while the 
other relations of Eq. (\ref{2.20}) are satisfied.

{\bf Solution II}

The nonlocal Hirota Eq. (\ref{2.1}) also admits a PT-invariant solution
with PT-eigenvalue +$1$, i.e.
\be\label{2.21}
u(x,t) = \sqrt{n}\, [A+iB\tanh(\xi)] e^{i(\omega t-kx)}\,,~~\xi = 
\delta(x-vt)\,,
\ee
provided
\bea\label{2.22}
&&g = -1\,,~~A^2+B^2 = 1\,,~~\delta = \sqrt{n} B\,,
~~\omega = -\alpha(k^2+2n)+\beta k(k^2+6n)\,, \nonumber \\
&&v = -2\alpha(k+\sqrt{n} A) +\beta(4n B^2-3k^2-6n-6kA\sqrt{n})\,.
\eea
Note that Eq. (\ref{2.22}) is also the solution of the 
corresponding local Hirota equation (\ref{2.1a}) with the same
constraints as given by Eq. (\ref{2.22}).
It is amusing to note that while the solution I holds good in case $g = 1$,
the solution II only holds good if $g = -1$.

{\bf Solution III}

It is not difficult to check that both the local (\ref{2.1a}) as well as the 
nonlocal Hirota Eq. (\ref{2.1}) admit the plane wave solution
\be\label{2.1d}
u(x,t) = A e^{i(\omega t-kx)}\,,
\ee
provided
\be\label{2.1e}
\omega = -\alpha(k^2-2g A^2) +\beta k(k^2-6gA^2)\,.
\ee

In order to obtain the other solutions of the nonlocal Hirota Eq. (\ref{2.1}),
we start with the ansatz
\be\label{2.2}
u(x,t) = e^{i\omega t} \phi(\xi)\,,~~\xi = \delta(x-vt)\,.
\ee
On substituting this ansatz in Eq. (\ref{2.1}) we obtain
\bea\label{2.3}
&&\alpha\left[\delta^2 \phi_{\xi \xi}(\xi)+2g \phi^2(\xi)\phi(-\xi) 
-\frac{\omega \phi}{\alpha}\right] \nonumber \\
&&+i\beta \delta \left[\delta^2 \phi_{\xi \xi \xi}(\xi) +6g \phi(\xi)
\phi(-\xi) \phi_{\xi}(\xi)- \frac{v \phi_{\xi}(\xi)}{\beta}\right]=0\,.
\eea
The two cases in which we can obtain the solutions of the 
Eq. (\ref{2.3}) are when $\phi(-\xi) = \pm \phi(\xi)$. This is because
in that case we can rewrite Eq. (\ref{2.3}) as 
\bea\label{2.4}
&&\alpha\left[\delta^2 \phi_{\xi \xi}(\xi)\pm 2g \phi^3(\xi) 
-\frac{\omega \phi}{\alpha}\right] \nonumber \\
&&+i\beta \delta \frac{d}{d\xi}\left[\delta^2 \phi_{\xi \xi}(\xi) \pm 2g \phi^3(\xi)
- \frac{v \phi(\xi)}{\beta}\right]=0\,,
\eea
so that as long as
\be\label{2.5}
\frac{\omega}{\alpha} = \frac{v}{\beta}\,,
\ee
then all solutions of equation
\be\label{2.6}
\delta^2 \phi_{\xi \xi}(\xi) = \frac{\omega}{\alpha} \phi \mp 2g \phi^3(\xi)\,,
\ee
automatically solve Eq. (\ref{2.4}) as long as the relation (\ref{2.5}) 
is valid. We now present several solutions of Eq. (\ref{2.6}) and hence
nonlocal Hirota Eq. (\ref{2.1}) satisfying relation (\ref{2.5}).

{\bf Solution IV}

It is easy to check that 
\be\label{2.9}
u(x,t) = e^{i\omega t} \frac{A\sqrt{m}\cn(\delta \xi,m)}
{D+\dn(\delta \xi,m)}\,,~~A, D > 0\,,
\ee
is an exact periodic pulse solution of the nonlocal Hirota Eq. (\ref{2.1}) 
provided relation (\ref{2.5}) is satisfied and further
\be\label{2.10}
D^2 = 1-m > 0\,,~~4g A^2 = -m \delta^2\,,~~\omega 
= -(2-m)\frac{\alpha \delta^2}{2}\,.
\ee
Note that this solution exists only if $g < 0, \omega < 0$. 
It is worth pointing out that Eq. (\ref{2.9}) is also the solution of the 
corresponding local Hirota equation (\ref{2.1a}) provided Eq. (\ref{2.10}) 
is satisfied.

{\bf Solution V}

It is easy to check that 
\be\label{2.11}
u(x,t) = e^{i\omega t}\left[F-\frac{A\dn(\delta \xi,m)}{D+\dn(\delta \xi,m)}\right]\,,
~~D,F,A > 0\,,
\ee
is an exact periodic pulse solution of the nonlocal Hirota Eq. (\ref{2.1}) 
provided relation (\ref{2.5}) is satisfied and further
\bea\label{2.12}
&&D^2 = \sqrt{1-m} > 0\,,~~ g A^2 = -(1-\sqrt{1-m})^2 \delta^2\,,
\nonumber \\
&&F = A/2\,,~~\omega = -[(2-m)+6\sqrt{1-m}]\frac{\delta^2}{2}\,.
\eea
Note that this solution exists only if $g < 0, \omega < 0$. 
It is worth pointing out that Eq. (\ref{2.11}) is also the solution of the 
corresponding local Hirota Eq. (\ref{2.1a}) provided the relations
(\ref{2.12}) are satisfied.

{\bf Solution VI}

A novel complex (but which is not PT-invariant) solution of the nonlocal
Hirota Eq. (\ref{2.1}) is 
\be\label{2.13}
u(x,t) = e^{i\omega t} \left[\frac{A\sqrt{1-m}}{\dn(\delta x,m)} 
+\frac{iB\sqrt{m}\cn(\delta x,m)}{\dn(\delta x,m)}\right]\,,
\ee
provided relation (\ref{2.5}) is satisfied and if further
\be\label{2.14}
B = \pm A\,,~~0 < m < 1\,,~~4g A^2 = -\delta^2\,,~~\omega 
= -(2m-1)\frac{\delta^2}{2}\,.
\ee
Note that Eq. (\ref{2.13}) is also the solution of the 
corresponding local Hirota Eq. (\ref{2.1a}) provided the relations
(\ref{2.14}) are satisfied.

{\bf Solution VII}

Another novel complex (but which is not PT-invariant) solution of the 
nonlocal Hirota Eq. (\ref{2.1}) is
\be\label{2.15}
u(x,t) = e^{i\omega t} \left[\frac{A\sqrt{m}\cn(\delta x,m)}{\dn(\delta x,m)} 
+\frac{iB\sqrt{1-m}}{\dn(\delta x,m)}\right]\,,
\ee
provided relation (\ref{2.5}) is satisfied and if further 
\be\label{2.16}
B = \pm A\,,~~0 < m < 1\,,~~4g A^2 = \delta^2\,,~~\omega
= -(2m-1)\frac{\delta^2}{2}\,.
\ee
Note that Eq. (\ref{2.15}) is also the solution of the 
corresponding local Hirota Eq. (\ref{2.1a}) provided the relations
(\ref{2.16}) are satisfied.

{\bf Solution VIII}

It is easy to check that 
\be\label{2.17}
u(x,t) = e^{i\omega t} \frac{A\sqrt{m}\sn(\delta \xi,m)}
{D+\dn(\delta \xi,m)}\,,~~A,D > 0\,,
\ee
is an exact periodic kink solution of the nonlocal Hirota Eq. (\ref{2.1}) 
provided relation (\ref{2.5}) is satisfied and if further
\be\label{2.18}
D = 1\,,~~4 g A^2 = \delta^2\,,~~\omega = -(2-m)\frac{\delta^2}{2}\,.
\ee
Note that Eq. (\ref{2.17}) is also the solution of the 
corresponding local Hirota Eq. (\ref{2.1a}) provided unlike the solution V,
$g < 0$ while the other relations of Eq. (\ref{2.18}) are satisfied.

{\bf Solution IX}

We now show that the nonlocal Hirota Eq. (\ref{2.1}) also admits solutions
with a more general ansatz of the form
\be\label{8.1}
u(x,t) = A e^{i(\omega t - kx)} \phi(\xi)\,,~~\xi = \delta(x-vt)\,.
\ee
In this case the relation (\ref{2.5}) between $\omega$ and $v$ is no 
more valid. We now present 12 such solutions. 

It is easy to check that 
\be\label{8.2}
u(x,t) = A e^{i(\omega t - kx)} \sqrt{m} \sn(\xi,m)\,,~~\xi = \delta(x-vt)\,,
\ee
is an exact periodic kink solution of Eq. (\ref{2.1}) provided
\bea\label{8.3}
&&g = 1\,,~~A = \delta\,,~~v = -2\alpha k -\beta(3k^2+(1+m)A^2)\,,
\nonumber \\
&&\omega = -\alpha[k^2+(1+m)A^2] +\beta k [k^2+3(1+m)A^2]\,.
\eea
Note that in the limit $k = 0$ we recover the solution obtained 
earlier \cite{ks2} satisfying relation (\ref{2.5}). 

{\bf Solution X}

In the limit $m = 1$, the solution VIII goes over to the hyperbolic 
moving kink solution
\be\label{8.4}
u(x,t) = A e^{i(\omega t - kx)} \tanh(\xi)\,,~~\xi = \delta(x-vt)\,,
\ee
provided
\bea\label{8.5}
&&g = 1\,,~~A = \delta\,,~~v = -2\alpha k -\beta(3k^2+2 A^2)\,,
\nonumber \\
&&\omega = -\alpha[k^2+2A^2] +\beta k [k^2+6 A^2]\,.
\eea
As expected, in the limit $k = 0$ we recover the solution obtained 
earlier \cite{ks2} satisfying relation (\ref{2.5}). 

{\bf Solution XI}

It is easy to check that 
\be\label{8.6}
u(x,t) = A e^{i(\omega t - kx)} \dn(\xi,m)\,,~~\xi = \delta(x-vt)\,,
\ee
is an exact periodic pulse solution of Eq. (\ref{2.1}) provided
\bea\label{8.7}
&&g = 1\,,~~A = \delta\,,~~v = -2\alpha k -\beta(3k^2-(2-m)A^2)\,,
\nonumber \\
&&\omega = -\alpha[k^2-(2-m)A^2] + \beta k [k^2-3(2-m)A^2]\,. 
\eea
Note that in the limit $k = 0$ we recover the solution obtained 
earlier \cite{ks2} satisfying relation (\ref{2.5}). 

{\bf Solution XII}

Yet another exact periodic pulse solution of nonlocal Hirota Eq.
(\ref{2.1}) is 
\be\label{8.8}
u(x,t) = A e^{i(\omega t - kx)} \sqrt{m} \cn(\xi,m)\,,~~\xi = \delta(x-vt)\,,
\ee
provided
\bea\label{8.9}
&&g = 1\,,~~A = \delta\,,~~v = -2\alpha k -\beta(3k^2-(2m-1)A^2)\,,
\nonumber \\
&&\omega = -\alpha[k^2-(2m-1)A^2]+ \beta k [k^2-3(2m-1)A^2]\,. 
\eea
Note that in the limit $k = 0$ we recover the solution obtained 
earlier \cite{ks2} satisfying relation (\ref{2.5}). 

{\bf Solution XIII}

In the limit $m = 1$, the solutions X and XI go over to the hyperbolic 
moving pulse solution
\be\label{8.10}
u(x,t) = A e^{i(\omega t - kx)} \sech(\xi)\,,~~\xi = \delta(x-vt)\,,
\ee
provided
\bea\label{8.11}
&&g = 1\,,~~A = \delta\,,~~v = -2\alpha k -\beta(3k^2- A^2)\,,
\nonumber \\
&&\omega = -\alpha[k^2-A^2] +\beta k [k^2-3 A^2]\,.
\eea
As expected, in the limit $k = 0$ we recover the solution obtained 
earlier \cite{ks2} satisfying relation (\ref{2.5}). 

{\bf Solution XIV}

It is easy to check that 
\be\label{8.12}
u(x,t) = e^{i(\omega t - kx)} \frac{A\sqrt{1-m}}{\dn(\xi,m)}\,,
~~\xi = \delta(x-vt)\,,
\ee
is an exact periodic pulse solution of Eq. (\ref{2.1}) provided
$0 < m < 1$ and further
\bea\label{8.13}
&&g = 1\,,~~A = \delta\,,~~v = -2\alpha k -\beta(3k^2-(2-m)A^2)\,,
\nonumber \\
&&\omega = -\alpha[k^2-(2-m)A^2]+ \beta k [k^2-3(2-m)A^2]\,. 
\eea
Note that in the limit $k = 0$ we recover the solution obtained 
earlier \cite{ks2} satisfying relation (\ref{2.5}). 

{\bf Solution XV}

Yet another exact periodic pulse solution of nonlocal Hirota Eq.
(\ref{2.1}) is 
\be\label{8.14}
u(x,t) =  e^{i(\omega t - kx)} \frac{A\sqrt{m(1-m)} \sn(\xi,m)}
{\dn(\xi,m)}\,,~~\xi = \delta(x-vt)\,,
\ee
provided $0 < m < 1$ and further
\bea\label{8.15}
&&g = -1\,,~~A = \delta\,,~~v = -2\alpha k -\beta(3k^2-(2m-1)A^2)\,,
\nonumber \\
&&\omega = -\alpha[k^2-(2m-1)A^2]+ \beta k[k^2-3(2m-1)A^2]\,.
\eea
Note that in the limit $k = 0$ we recover the solution obtained 
earlier \cite{ks2} satisfying relation (\ref{2.5}). 

{\bf Solution XVI}

It is easy to check that 
\be\label{8.16}
u(x,t) = e^{i(\omega t - kx)} \frac{A\sqrt{m}\cn(\xi,m)}{\dn(\xi,m)}\,,
~~\xi = \delta(x-vt)\,,
\ee
is an exact periodic kink solution of Eq. (\ref{2.1}) provided
$0 < m < 1$ and further
\bea\label{8.17}
&&g = -1\,,~~A = \delta\,,~~v = -2\alpha k -\beta(3k^2+(1+m)A^2)\,,
\nonumber \\
&&\omega = -\alpha[k^2+(1+m)A^2] +\beta k [k^2+3(1+m)A^2]\,. 
\eea
Note that in the limit $k = 0$ we recover the solution obtained 
earlier \cite{ks2} satisfying relation (\ref{2.5}). 

{\bf Solution XVII}

Remarkably, the nonlocal Hirota Eq. (\ref{2.1}) also admits the
superposed solution
\be\label{8.18}
u(x,t) = e^{i(\omega t - kx)} [A \dn(\xi,m)+ B\sqrt{m}\cn(\xi,m)]\,,
~~\xi = \delta(x-vt)\,,
\ee
provided $0 < m < 1$ and further
\bea\label{8.19}
&&g = 1\,,~~2A = \delta\,,~~B = \pm A\,,~~v = -2\alpha k 
-\beta[3k^2-2(1+m)A^2]\,, \nonumber \\
&&\omega = -\alpha[k^2-2(1+m)A^2] -\beta k [k^2-6(1+m)A^2]\,.
\eea
Note that in the limit $k = 0$ we recover the solution obtained 
earlier \cite{ks2} satisfying relation (\ref{2.5}). 

{\bf Solution XVIII}

Remarkably, the nonlocal Hirota Eq. (\ref{2.1}) also admits the
superposed solution
\be\label{8.20}
u(x,t) = e^{i(\omega t - kx)} \left[A \dn(\xi,m)+ \frac{B\sqrt{1-m}}{\dn(\xi,m)}\right]\,,
~~\xi = \delta(x-vt)\,,
\ee
provided $0 < m < 1$ and further
\bea\label{8.21}
&&g = 1\,,~~A = \delta\,,~~B = \pm A\,, \nonumber \\
&&v = -2\alpha k -\beta[3k^2-(2-mm)A^2-6\sqrt{1-m}A^2]\,, \nonumber \\
&&\omega = -\alpha[k^2-(2-m)A^2 -6\sqrt{1-m} A^2]  \nonumber \\
&&-\beta k[k^2 -3(2-m)A^2 -18\sqrt{1-m}A^2]\,.
\eea
Note that in the limit $k = 0$ we recover the solution obtained 
earlier \cite{ks2} satisfying relation (\ref{2.5}). 

{\bf Solution XIX}

Remarkably, the nonlocal Hirota Eq. (\ref{2.1}) also admits the
complex superposed solution
\be\label{8.22}
u(x,t) = e^{i(\omega t - kx)} \left[\frac{A\sqrt{1-m}}{\dn(\xi,m)}
+ \frac{iB\sqrt{m}\cn(\xi,m)}{\dn(\xi,m)}\right]\,,~~\xi = \delta(x-vt)\,,
\ee
provided $0 < m < 1$ and further
\bea\label{8.23}
&&g = 1\,,~~2A = \delta\,,~~B = \pm A\,,~~v = -2\alpha k 
-\beta[3k^2+2(2m-1)A^2]\,, \nonumber \\
&&\omega = -\alpha[k^2+2(2m-1)A^2] -k\beta[k^2+6(2m-1)A^2] \,.
\eea
Note that in the limit $k = 0$ we recover the solution V 
satisfying relation (\ref{2.5}). 

{\bf Solution XX}

The nonlocal Hirota Eq. (\ref{2.1}) also admits another
complex superposed solution
\be\label{8.24}
u(x,t) = e^{i(\omega t - kx)} \left[\frac{A\sqrt{m}\cn(\xi,m)}{\dn(\xi,m)}
+ \frac{iB\sqrt{1-m}}{\dn(\xi,m)}\right]\,,~~\xi = \delta(x-vt)\,,
\ee
provided $0 < m < 1$ and further
\bea\label{8.25}
&&g = -1\,,~~2A = \delta\,,~~B = \pm A\,,~~v = -2\alpha k 
-\beta[3k^2+2(2m-1)A^2]\,, \nonumber \\
&&\omega = -\alpha[k^2+2(2m-1)A^2] -k\beta[k^2+6(2m-1)A^2] \,.
\eea
Note that in the limit $k = 0$ we recover the solution VI
satisfying relation (\ref{2.5}). 

In Appendix C we present 5 solutions of the local Hirota Eq. (\ref{2.1a})
which however are not the solutions of the nonlocal Hirota Eq. (\ref{2.1}). 

\section{Conclusion and Open Problems}

In this paper we have obtained several new solutions of 
Ablowitz-Musslimani as well as Yang nonlocal variants of the NLS equations, 
nonlocal mKdV equation and nonlocal Hirota equation. Further, we have 
compared and contrasted with the solutions of the corresponding local NLS,
local mKdV and local Hirota equations, respectively. In particular, we found
that unlike the local mKdV equation, the nonlocal mKdV equation admits
not only a plane wave solution of the form $e^{i(kx-\omega t)}$, but also 
several solutions multiplied by the same factor. Besides, we found that
unlike the local NLS, the Ablowitz-Musslimani nonlocal variant of the NLS 
equation admits several complex PT-invariant solutions. For completeness,
in 3 appendices we have also mentioned those solutions of the local NLS, 
mKdV and Hirota equations which are, however, not the solutions of the 
corresponding nonlocal NLS, nonlocal mKdV and nonlocal Hirota equations.

There are several open questions. For example, there are many 
discrete nonlocal nonlinear equations like nonlocal Ablowitz-Ladik equation,
nonlocal discrete NLS (DNLS) equation or discrete saturable nonlocal DNLS
equation and the obvious question is if these nonlocal equations also admit 
similar new novel solutions. Secondly, whether the corresponding coupled nonlocal
equations like nonlocal coupled NLS and coupled nonlocal mKdV also admit
similar new solutions. We hope to address some of these issues in near future.

\section{Acknowledgment} 

One of us (AK) is grateful to Indian National Science Academy (INSA) for the award of INSA Honorary Scientist position at Savitribai Phule Pune University. The work at Los Alamos National Laboratory was carried out under the auspices of the US DOE and NNSA under contract No.~DEAC52-06NA25396.

\section{Appendix A: NLS Solutions Which are Not Solutions of the Nonlocal NLS Equation}

We now present six new solutions of the local NLS Eq. (\ref{1a}) which 
though are not the solutions of the Ablowitz-Musslimani nonlocal Eq. (\ref{1}).

{\bf Solution IA}

It is easy to check that
\be\label{A1}
\psi = e^{i\omega t} \frac{A\dn(\beta x,m)}{D+\sn(\beta x,m)}\,,
~~A > 0, D > 1\,,
\ee
is an exact periodic pulse solution of the local NLS Eq. (\ref{1}) provided
$0 < m < 1$ and further
\be\label{A2}
m D^2 = 1\,,~~4 g m A^2 = (1-m) \beta^2\,,~~\omega 
= (1+m)\frac{\beta^2}{2}\,.
\ee

{\bf Solution IIA}

It is easy to check that 
\be\label{A3}
\psi = e^{i\omega t} \frac{[A\dn(\beta x,m)+B\sqrt{m}\cn(\beta x,m)]}
{D+\sn(\beta x,m)}\,,~~A, B > 0, D > 1\,,
\ee
is an exact superposed periodic solution of the local NLS Eq. (\ref{1}) 
provided
\be\label{A4}
4 g A^2 = (D^2-1)\beta^2\,,~~4 m g B^2 = (m D^2-1)\beta^2\,,~~
\omega = (1+m)\frac{\beta^2}{2}\,.
\ee
Thus, this solution exists only if either $D^2 > 1/m, g > 0$ or $D^2 < 1,
g < 0$.

{\bf Solution IIIA}

It is easy to check that 
\be\label{A5}
\psi = e^{i\omega t} \left[F+\frac{A\sn(\beta x,m)}{D+\sn(\beta x,m)}\right]\,,
~~D > 1,F,A > 0\,,
\ee
is an exact periodic pulse solution of the local NLS Eq. (\ref{1}) provided
\bea\label{A6}
&&A = \frac{2 F}{\sqrt{m}}\,,~~g A^2 = -\frac{(1-\sqrt{m})^2 \beta^2}
{\sqrt{m}}\,,~~D^2 = \frac{1}{\sqrt{m}}\,,  \nonumber \\
&&0 < m < 1\,,~~\omega = -\left[\frac{3(3+m)\sqrt{m}}{2}+(2m-1)\right]\beta^2\,.
\eea

{\bf Solution IVA}

It is well known that the local NLS Eq. (\ref{1}) admits both 
$A\dn$ and $B\sn/\dn$ as exact solutions \cite{ks2} but we now show that 
remarkably, even their superposition is an exact solution of Eq. (\ref{1}). In 
particular, it is easy to check that
\be\label{A7}
\psi = e^{i\omega t} \left[A\dn(\beta x,m) +\frac{B\sqrt{m(1-m)}\sn(\beta x,m)}
{\dn(\beta x,m)}\right]\,,
\ee
is an exact solution of Eq. (\ref{1}) provided
\be\label{A8}
0 < m < 1\,,~~B = \pm A\,,~~4g A^2 = \beta^2\,,~~
\omega = (1+m)\beta^2\,.
\ee
Note, this solution is not an eigenstate of parity, i.e. while the first term
is even under $x \rightarrow -x$, the second term is odd under parity.

{\bf Solution VA}

It is not so well known that the local NLS Eq. (\ref{1a}) admits the 
PT-invariant solution with PT-eigenvalue $-1$, i.e. 
\be\label{A9}
\psi = \sqrt{n}[B\tanh(\xi)+iA] e^{i(kx-\omega t + \theta_0)}\,,~~\xi 
= \beta(x-vt+t_0)\,,
\ee
provided
\be\label{A10}
g = -1\,,~~A^2+B^2 = 1\,,~~\beta = \sqrt{n/2} B\,,~~\omega = k^2+n\,,~~
v = 2k +\sqrt{2n} A\,.
\ee

{\bf Solution VIA}

The local NLS Eq. (\ref{1a}) also admits the PT-invariant solution with
PT-eigenvalue +$1$, i.e.
solution
\be\label{A11}
\psi = \sqrt{n}[A+iB\tanh(\xi)] e^{i(kx-\omega t + \theta_0)}\,,~~\xi 
= \beta(x-vt+t_0)\,,
\ee
provided
\be\label{A12}
g = -1\,,~~A^2+B^2 = 1\,,~~\beta = \sqrt{n/2} B\,,~~\omega = k^2+n\,,~~
v = 2k -\sqrt{2n} A\,.
\ee
Notice that the solutions VA and VIA hold good under the same conditions
except for the relation for the velocity $v$.

\section{Appendix B: mKdV Solutions Which are Not Solutions of the Nonlocal mKdV Equation}

We now present 14 new solutions of the (local) mKdV equation (\ref{1.1a}),
neither of which is a solution of the nonlocal mKdV Eq. (\ref{1.1}).

{\bf Solution IB}

It is easy to check that
\be\label{B1}
u(\xi) = F - \frac{A}{D+\sn(\xi,m)}\,,~~D > 1\,,~~\xi = \beta(x-vt)\,,
\ee
is an exact solution of mKdV Eq. (\ref{2.1a}) provided
\bea\label{B2}
&&~~v -6g F^2 = [6m D^2-(1+m)]\beta^2\,, \nonumber \\
&&2gFA = D[(1+m)-2m D^2]\beta^2\,,~~gA^2 = -(D^2-1)(m D^2-1) \beta^2\,. ~~~~~~
\eea
Since $D > 1$, it implies that
\bea\label{B3}
&&1 < D^2 < 1/m\,,~~g = 1\,, \nonumber \\
&&D^2 > 1/m\,,~~g  = -1\,.
\eea
This solution can be re-expressed as
\be\label{B4}
u(\xi) = \frac{FD-A+F\sn(\xi,m)}{D+\sn(\xi,m)}\,,~~D > 1\,,
~~\xi = \beta(x-vt)\,,
\ee
There are two special cases in which the Solution IB takes a simpler
form which we now discuss one by one.

{\bf Case a: $F = 0$}

In case $F = 0$, the solution IB takes the simpler form
\be\label{B5}
u(\xi) = -\frac{A}{D+\sn(\xi,m)}\,,~~D > 1\,,~~\xi = \beta(x-vt)\,,
\ee
provided
\be\label{B6}
g = -1\,,~~v = 2(1+m)\beta^2\,,~~D^2 = \frac{(1+m)}{2m}\,,~~
4 A^2 = (1-m)^2 \beta^2\,.
\ee

{\bf Case b: $FD = A$}

In case $FD = A$, the solution IB takes the simpler form
\be\label{B7}
u(\xi) = \frac{F\sn(\xi,m)}{D+\sn(\xi,m)}\,,~~D > 1\,,~~\xi = \beta(x-vt)\,,
\ee
provided
\be\label{B8}
g = 1\,,~~v = 2(1+m)\beta^2\,,~~D^2 = \frac{2}{(1+m)}\,,~~
F^2 = \frac{(1-m)^2 \beta^2}{2(1+m)}\,.
\ee

{\bf Solution IIB}

In the limit $m = 1$, the solution  IB goes over to the hyperbolic 
kink solution
\be\label{B9}
u(\xi) = F - \frac{A}{D+\tanh(\xi)}\,,~~D > 1\,,
\ee
provided
\be\label{B10}
g = -1\,,~~v+6 F^2 = 2[3D^2-1]\beta^2\,,~~ FA = D(D^2-1)\beta^2\,,
A^2 = (D^2-1)^2 \beta^2\,.
\ee

{\bf Solution IIIB}

It is easy to check that
\be\label{B8a}
u(\xi) = \frac{A\dn(\beta x,m)}{D+\sn(\beta \xi,m)}\,,~~D > 1\,,
~~\xi = \beta(x-vt)\,,
\ee
is an exact solution of mKdV Eq. (\ref{2.1a}) provided
\be\label{B9a}
g = 1\,,~~v = \frac{(1+m)\beta^2}{2}\,,~~D = m^{-1/2}\,,
~~A^2 = \frac{(1-m)\beta^2}{4m}\,.
\ee

{\bf Solution IVB}

It is easy to check that
\be\label{B10a}
u(\xi) = \frac{A\dn(\beta x,m)+B\sqrt{m}\cn(\beta x,m)}{D+\sn(\beta \xi,m)}\,,
~~D > 1\,,~~\xi = \beta(x-vt)\,,
\ee
is an exact solution of mKdV Eq. (\ref{2.1a}) provided
\be\label{B11}
g = 1\,,~~v = \frac{(1+m)\beta^2}{2}\,,~~4A^2 = (D^2-1)\beta^2\,,~~
4m B^2 = (mD^2-1)\beta^2\,.
\ee
Thus this solution exists only if $m D^2 > 1$.

{\bf Solution VB}

Remarkably, mKdV Eq. (\ref{1.1a}) also admits a complex PT-invariant 
pulse solution with PT-eigenvalue +$1$
\be\label{B12}
u(\xi) = \frac{[A\sqrt{m}\cn(\beta \xi,m)+iB\sqrt{m} \sn(\beta \xi,m)]}
{D+\dn(\beta \xi,m)}\,,~~D > 0\,,
\ee
provided
\be\label{B13}
4g A^2 = (D^2-1)\beta^2\,,~~4g B^2 = (D^2-1+m)\beta^2\,,~~
v = -\frac{(2-m)\beta^2}{2}\,.
\ee
Thus there are two possibilities.

1. $g = 1\,,~~m D^2 > 1$. 

2. $g = -1\,,~~D^2 < 1-m$. Thus in this case $4A^2 = (1-D^2)\beta^2\,,
~~4B^2 = (1-m-D^2)\beta^2$.

{\bf Solution VIB}

Remarkably, mKdV Eq. (\ref{1.1a}) also admits another complex PT-invariant 
pulse solution with PT-eigenvalue +$1$
\be\label{B14}
u(\xi) = \frac{[A\dn(\beta \xi,m)+iB\sqrt{m} \sn(\beta \xi,m)]}
{D+\cn(\beta \xi,m)}\,,~~D > 1\,,
\ee
provided
\be\label{B15}
g=1\,,~~4 A^2 = (D^2-1)\beta^2\,,~~4m B^2 = (m D^2 -m+1) \beta^2\,,~~
v = -\frac{(2m-1)\beta^2}{2}\,.
\ee
Thus this solution only exists if $v < 0$ and $D^2 > 1$. 

{\bf Solution VIIB}

In the limit $m =1$, the solutions VI and VII go over to the hyperbolic 
complex PT-invariant pulse solution with PT-eigenvalue +$1$
\be\label{B16}
u(\xi) = \frac{[A\sech(\beta \xi)+iB \tanh(\beta \xi)]}
{D+\sech(\beta \xi)}\,,~~D > 0\,,
\ee
provided
\be\label{B17}
g=1\,,~~4 A^2 = (D^2-1)\beta^2\,,~~4 B^2 = D^2 \beta^2\,,~~
v = -\frac{\beta^2}{2}\,.
\ee
Thus this solution only exists if $v < 0$ and $D^2 > 1$. 

{\bf Solution VIIIB}

Remarkably, mKdV Eq. (\ref{1.1a}) also admits a complex PT-invariant 
kink solution with PT-eigenvalue $-1$
\be\label{B18}
u(\xi) = \frac{[A\sqrt{m} \sn(\beta \xi,m)+iB\sqrt{m} \cn(\beta \xi,m)]}
{D+\dn(\beta \xi,m)}\,,~~D > 0\,,
\ee
provided
\be\label{B19}
4g A^2 = -(D^2-1+m) \beta^2\,,~~4g B^2 = -(D^2-1)\beta^2\,,~~
v = -\frac{(2-m)\beta^2}{2}\,.
\ee
Thus there are two possibilities.

1. $g = -1\,,~~ D^2 > 1$. 

2. $g = 1\,,~~D^2 < 1-m$. Thus in this case $4B^2 = (1-D^2)\beta^2\,,
~~4A^2 = (1-m-D^2)\beta^2$.

{\bf Solution IXB}

Remarkably, mKdV Eq. (\ref{1.1a}) also admits another complex PT-invariant 
kink solution with PT-eigenvalue $-1$
\be\label{B20}
u(\xi) = \frac{[A\sqrt{m} \sn(\beta \xi,m)+iB \dn(\beta \xi,m)]}
{D+\cn(\beta \xi,m)}\,,~~D > 1\,,
\ee
provided
\be\label{B21}
g = -1\,,~~4m A^2 = (mD^2+1-m) \beta^2\,,~~4 B^2 = (D^2-1)\beta^2\,,~~
v = -\frac{(2m-1)\beta^2}{2}\,.
\ee

{\bf Solution XB}

In the limit $m = 1$, both the periodic complex kink solutions IX and X
go over to the hyperbolic complex kink solution with PT-eigenvalue $-1$
\be\label{B22}
u(\xi) = \frac{[A\tanh(\beta \xi)+iB\sech(\beta \xi)]}{D+\sech(\beta \xi)}\,,
~~D > 0\,,
\ee
provided
\be\label{B23}
g=-1\,,~~4 A^2 = D^2 \beta^2\,,~~4 B^2 = (D^2-1)\beta^2\,,~~
v = -\frac{\beta^2}{2}\,.
\ee
Thus this solution only exists if $v < 0$ and $D^2 > 1$. 

{\bf Solution XIB}

It is easy to check that
\be\label{B24}
u(\xi) = \frac{A\sin(\beta \xi)+iB}{D+\cos(\beta \xi)}\,,~~D > 1\,,
\ee
is an exact solution of mKdV Eq. (\ref{1.1a}) provided 
\be\label{B25}
g = -1\,,~~v = \frac{\beta^2}{2} > 0\,,~~4B^2 =-(D^2-1)\beta^2\,,
4 A^2 = \beta^2\,.
\ee

{\bf Solution XIIB}

It is straightforward to check that
\be\label{B26}
u(\xi) = \frac{A+iB\sin(\beta \xi)}{D+\cos(\beta \xi)}\,,~~D > 1\,,
\ee
is an exact solution of mKdV Eq. (\ref{1.1a}) provided 
\be\label{B27}
g = 1\,,~~v = \frac{\beta^2}{2} > 0\,,~~4 A^2 =-(D^2-1)\beta^2\,,
4 B^2 = \beta^2\,.
\ee

{\bf Solution XIIIB}

It is not difficult to check that
\be\label{B28}
u(\xi) = \frac{\sqrt{1-m} A+B \sqrt{m(1-m)}\sn(\beta \xi,m)}
{\dn(\beta \xi, m)}\,,
\ee
is an exact solution of mKdV Eq. (\ref{1.1a}) provided 
\be\label{B29}
g = -1\,,~~v = \frac{(1+m)\beta^2}{2} > 0\,,~~4 A^2 = \beta^2\,.
\ee

Finally, we want to remind about the celebrated Miura transformation \cite{miura} which 
showed a remarkable connection between the solutions of the repulsive mKdV 
(i.e. Eq. (\ref{1.1a}) with $g = -1$) and the KdV equation
\be\label{B30}
w_t + w_{xxx} - 6 w w_{x} = 0\,.
\ee
In particular it was shown that if $u(x,t)$ is a solution of the repulsive
mKdV equation
\be\label{B31}
u_t + u_{xxx} - 6 u^2 u_{x} = 0\,,
\ee
then $w(x,t) = u^2(x,t)+u_x(x,t)$ is the corresponding solution of the 
KdV Eq. (\ref{B30}). Out of the 14 solutions given above, we find that the
solutions IB, IIIB, XB, XIB, XIIB, XIVB  hold good when $g = -1$ while 
solutions IIB, VIB, IXB hold good for $g = -1$ under certain conditions. 
Besides, solutions IV, V and VI of Sec. IV  also hold good in the case of local 
mKdV provided $g = -1$. Besides, solutions VII and IX hold good for $g = -1$ 
in the case of local mKdV under certain conditions. Thus in all these cases, 
the local KdV Eq. (\ref{B30}) admits solutions related to the corresponding
mKdV solutions by the Miura transformation. 

\section{Appendix C: Solutions of Local Hirota equation Which are Not Solutions of 
the Nonlocal Hirota Equation}

We now present 5 new solutions of (local) Hirota Eq. (\ref{2.1a}) which 
are, however, not the solutions of the nonlocal Hirota Eq. (\ref{2.1}).

{\bf Solution IC}

It is easy to check that 
\be\label{C1}
u(x,t) = e^{i\omega t} \frac{A\dn(\delta \xi,m)}{D+\sn(\delta \xi,m)}\,,
~~D > 1\,,~~\xi = x-vt\,,
\ee
is an exact periodic pulse solution of (local) Hirota Eq. (\ref{2.1a}) 
provided
\bea\label{C2}
&&0 < m < 1\,,~~ m D^2 = 1\,,~~4g m A^2 = (1-m) \delta^2\,,
\nonumber \\
&&\frac{\omega}{\alpha} = (1+m)\frac{\delta^2}{2}\,,
~~\omega \beta = v \alpha\,.
\eea

{\bf Solution IIC}

It is straightforward to check that 
\be\label{C3}
u(x,t) = e^{i\omega t} \frac{[A\dn(\delta \xi,m)+\sqrt{m}\cn(\delta \xi,m)]}
{D+\sn(\beta x,m)}\,,~~D > 1\,,~~\xi = x-vt\,,
\ee
is an exact superposed periodic solution of Hirota Eq. (\ref{2.1a}) provided
\bea\label{C4}
&&4g A^2 = (D^2-1)\delta^2\,,~~4 m g B^2 = (m D^2-1)\delta^2\,,
\nonumber \\
&&\frac{\omega}{\alpha} = (1+m)\frac{\delta^2}{2}\,,
~~\omega \beta = v \alpha\,.
\eea
Thus such a solution exists only if $m D^2 > 1$. 

{\bf Solution IIIC}

In the limit $m = 1$, the solution II does not exist but instead we have a 
novel pulse solution
\be\label{C5}
u(x,t) = e^{i\omega t} \frac{A\sech(\delta \xi)}{D+\tanh(\delta \xi)}\,,
~~D > 1\,,~~\xi = x -vt\,,
\ee
provided
\be\label{C6}
g A^2 = (D^2-1) \delta^2\,,~~\frac{\omega}{\alpha} = \delta^2\,, 
~~\omega \beta = v \alpha\,.
\ee

{\bf Solution IVC}

It is easy to check that 
\be\label{C7}
u(x,t) = e^{i\omega t} \left[F+\frac{A\sn(\delta \xi,m)}{D+\sn(\delta \xi,m)}\right]\,,
~~A,F> 0\,,~~ D > 1\,,~~\xi = x-vt\,,
\ee
is an exact periodic pulse solution of Hirota Eq. (\ref{2.1a}) provided 
\bea\label{C8}
&&0 < m < 1\,,~~ g A^2 = -\frac{(1-\sqrt{m})^2 \delta^2}{\sqrt{m}}\,,
~~F = A/2\,, \nonumber \\
&&\frac{\omega}{\alpha} = -[3(3+m)\sqrt{m}+2(2m-1)]\frac{\delta^2}{2}\,, 
~~\omega \beta = v \alpha\,.
\eea

{\bf Solution VC}

It is easy to check that 
\be\label{C9}
u(x,t) = e^{i\omega t} \frac{[A\sqrt{1-m}+B\sqrt{m(1-m)}\sn(\delta \xi,m)]}
{\dn(\beta x,m)}\,,~~\xi = x-vt\,,
\ee
is an exact superposed periodic solution of Hirota Eq. (\ref{2.1a}) provided
\bea\label{C10}
&&B = \pm A\,,~~4g A^2 = \delta^2\,,~~\omega \beta = v \alpha\,,
\nonumber \\
&&\frac{\omega}{\alpha} = (1+m)\frac{\delta^2}{2}\,.
\eea





\begin{thebibliography}{99}

\bibitem{am1} M. J. Ablowitz and Z. H. Musslimani, Phys. Rev. Lett. 
{\bf 110} (2013) 064105.

\bibitem{am2} M. J. Ablowitz and Z. H. Musslimani, Nonlinearity {\bf 29}
(2016) 915.

\bibitem{mkdv} F. He, E. Fan and J. Xu, arXiv:1804.10863.

\bibitem{gur} M. G\"urses and A. Peckan, arXiv:1711.01588; Comm. Nonlin.
Sc. Num. Simul. {\bf 67} (2019) 427.

\bibitem{cen} J. Cen, F. Correa and A. Fring, J. Math. Phys. {\bf 60}
(2019) 081508.

\bibitem{hirota} Y. Zia, R. Yao and X. Xin, Chinese Physics {\bf B31} 
(2021) 020401

\bibitem{yang} J. Yang, Phys. Rev. {\bf E98} (2018) 042202.

\bibitem{optics} Z. H. Musslimani et al., Phys. Rev. Lett. {\bf 100}
(2008) 030402; K. G. Makris et al., ibid. {\bf 100} (2008) 103904; 
A. Ruschapupt, F. Delgado and J. G. Mugga, J. Phys. {\bf A38} (2005), 
L171; A. Guo et al., Phys. Rev. Lett. {\bf 103} (2009) 093902; C. E.
Rutter, K. G. Makris, R. El-Ganainy, R. N. Christodulides, M. Segev 
and D. Kip, Nature Phys. {\bf 6} (2010) 192. 

\bibitem{lou} S. Y. Lou, Sci.  Rep. {\bf 7} (2017) 869. 

\bibitem{ks1} A. Khare and A. Saxena, J. Math. Phys. {\bf 55} (2014) 
032701.

\bibitem{ks3} A. Khare and A. Saxena, J. Math. Phys. {\bf 56} (2015) 
032104.


\bibitem{am3} M. J. Ablowitz and Z.H. Musslimani, Nonlinearity {\bf 29} 
(2016) 915.

\bibitem{am4} M. J. Ablowitz, X. D. Luo and Z.H. Musslimani, J. Math. Ph.
{\bf 59} (2018) 011501; Nonlinearity {\bf 31} (2018) 5385.

\bibitem{li} J. L. Li and Z. N. Zhu, J. Math. Anal. Appl. {\bf 453} 
(2017) 7.
\bibitem{zh} G. Q. Zhang and Z.Y. Yan, Physica {\bf D402} (2020) 132170

\bibitem{li1} Y. Li and R. Guo, Nonlin. Dyn. {\bf 105} (2021) 617 

\bibitem{ks2} A. Khare and A. Saxena, J. Math. Phys. {\bf 63} (2022) 122903.

\bibitem{kbs} A. Khare, S. Banerjee and A. Saxena, Ann. of Phys. 
{\bf 452} (2023) in press; arXiv:2303.03737. 

\bibitem{as} See, for example, M. Abramowitz and I. Stegun, {\it Handbook of 
Mathematical Functions with Formulas, Graphs and Mathematical Tables}, Dover,
NY (1964).

\bibitem{zs} V. S. Zakharov and A. B. Shabat, JETP {\bf 37} (1973) 823.

\bibitem{miura} R. Miura, J. Math. Phys. {\bf 9}, 1202 (1968). 

\end{thebibliography}
\end{document}